\documentclass[prl,twocolumn,preprintnumbers,amsmath,amssymb,superscriptaddress]{revtex4} 
\usepackage[utf8]{inputenc}
\usepackage{epsfig}
\usepackage{subfigure}
\usepackage{bm}
\usepackage{braket}
\usepackage[dvipsnames]{xcolor}
\usepackage{times}
\usepackage{amsmath,amsthm,amssymb}
\usepackage[colorlinks,bookmarks=false,citecolor=blue,linkcolor=red,urlcolor=blue]{hyperref}

\usepackage{soul}

\newcommand{\be}{\begin{equation}}
\newcommand{\ee}{\end{equation}}
\newcommand{\bea}{\begin{eqnarray}}
\newcommand{\eea}{\end{eqnarray}}

\def\Tr{\mathop{\mbox{\normalfont Tr}}\nolimits}

\begin{document}
\setstcolor{red}
\title{Entanglement asymmetry as a probe of symmetry breaking
}
\author{Filiberto Ares}
\email{faresase@sissa.it}
\affiliation{SISSA and INFN, via Bonomea 265, 34136 Trieste, Italy}
 
\author{Sara Murciano}
\affiliation{SISSA and INFN, via Bonomea 265, 34136 Trieste, Italy}
\affiliation{Department of Physics and Institute for Quantum Information and Matter, California Institute of Technology, Pasadena, CA 91125, USA}
\affiliation{Walter Burke Institute for Theoretical Physics, California Institute of Technology, Pasadena, CA 91125, USA}

 \author{Pasquale Calabrese}
 \affiliation{SISSA and INFN, via Bonomea 265, 34136 Trieste, Italy}
 \affiliation{The Abdus Salam International Center for Theoretical Physics, Strada Costiera 11, 34151 Trieste, Italy}

\maketitle

{\bf 
Symmetry and symmetry breaking are two pillars of modern quantum physics. 
Still, quantifying how much a symmetry is broken is an issue that has received little attention. 
In extended quantum systems, this problem is intrinsically bound to the subsystem of interest. 
Hence, in this work, we borrow methods from the theory of entanglement in many-body quantum systems
to introduce a subsystem measure of symmetry breaking that we dub {\emph{ entanglement asymmetry}}.  
As a prototypical illustration, we study the  entanglement asymmetry  in a quantum quench of a spin chain in which an initially broken global $U(1)$ symmetry is restored dynamically. 
We adapt the quasiparticle picture for entanglement evolution to the analytic determination of the entanglement asymmetry. 
We find, expectedly, that larger is the subsystem, slower is the restoration, but also the counterintuitive result that more the symmetry is initially broken, faster it is restored, 
a sort of quantum Mpemba effect, a phenomenon that we show to occur in a large variety of systems.} 

{\bf Introduction}\\
Symmetries hold a special place in every branch of physics, from relativity to quantum mechanics, passing through gauge/gravity duality and numerical algorithms. It is difficult to identify who was the first in understanding their relevance since the transversal development of the subject is a huge puzzle where different scientists, from Galileo to Noether, gave their own remarkable contributions. 
Sometimes it happens that, when a parameter reaches a critical value, the lowest energy configuration respecting the symmetry of the theory becomes unstable  
and new asymmetric lowest energy solutions can be found. 
This phenomenon does not require an input, whence the name spontaneous symmetry breaking. 
Other times a symmetry can be explicitly broken, in the sense that the Hamiltonian describing the system contains terms that manifestly break it. 
The present work fits in this framework: our main goal is to find a tool that measures quantitatively how much a symmetry is broken.

To be more specific, the setup we are interested in is an extended quantum system in a pure state $\ket{\Psi}$, 
which we divide into two spatial regions $A$ and $B$. The state 
of $A$ is described by the reduced density matrix 
$\rho_A=\mathrm{Tr}_B(\ket{\Psi}\bra{\Psi})$. We consider
a charge operator $Q$ 
that generates a global $U(1)$ symmetry group, hence satisfying 
$Q=Q_A+Q_B$. 
If $\ket{\Psi}$ is an eigenstate of $Q$, then $[\rho_A, Q_A]=0$ and $\rho_A$ displays a block-diagonal structure, with each block corresponding to 
a charge sector of $Q_A$. Thus the entanglement entropy $S(\rho_A)=-\mathrm{Tr}(\rho_A\log\rho_A)$,
which measures how entangled $A$ and $B$ are, can be decomposed
into the contributions of each charge sector~\cite{lr-14, goldstein, xavier,MDC-20,riccarda,mcp-22} (known as  symmetry-resolved
entanglement), recently accessed also experimentally~\cite{fis, Azses, Neven, Vitale}. 

\begin{figure}[b]
     \centering
     \includegraphics[width=\linewidth]{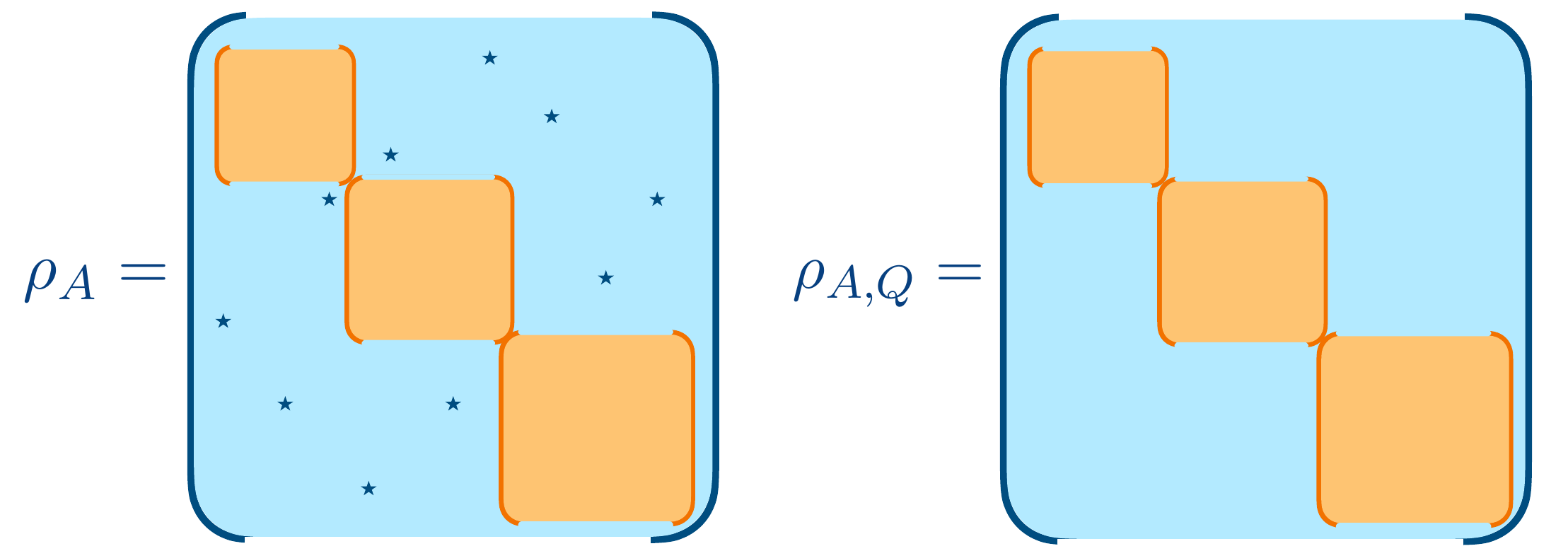}
     \caption{{\bf The density matrices $\rho_A$ and $\rho_{A, Q}$ entering in the definitions \eqref{eq:def} and \eqref{eq:replicatrick} of the entanglement asymmetries.}
     In the eigenbasis of the subsystem charge $Q_A$, $\rho_A$ generically displays off-diagonal elements. 
     Under a projective measurement of $Q_A$, we get $\rho_{A,Q}$, where the off-diagonal blocks are annihilated. 
     The difference $\Delta S_{A}^{(n)}$ between the entanglement entropies of these matrices is our probe of symmetry breaking.}
     \label{fig:cartoon}
\end{figure}

Here we consider the opposite situation: a 
state $\ket{\Psi}$ that breaks the global $U(1)$ symmetry. Therefore, $[\rho_A, Q_A]\neq 0$ and $\rho_A$ is not block-diagonal in the eigenbasis of $Q_A$. 
The goal of this work is to introduce a quantifier of the  symmetry breaking at the level of the subsystem $A$, 
which is the {\it entanglement asymmetry} defined as
\begin{equation}\label{eq:def}
 \Delta S_{A}=S(\rho_{A,Q})-S(\rho_A).
\end{equation}
Here $\rho_{A,Q}=\sum_{q\in\mathbb{Z}}\Pi_q\rho_A\Pi_q$, where $\Pi_q$ is the projector onto the eigenspace of $Q_A$ with charge $q\in\mathbb{Z}$. 
Thus $\rho_{A, Q}$ is block-diagonal in the eigenbasis of $Q_A$. 
In Fig.~\ref{fig:cartoon}, we pictorially show how $\rho_{A, Q}$ is obtained from $\rho_{A}$. 
A similar quantity, but  for the full system, has been recently introduced in Ref.~\cite{ms-21} to study the inseparability of mixed states with 
a globally conserved charge.

The entanglement asymmetry~\eqref{eq:def} satisfies two natural properties to quantify symmetry breaking:
$(i)$ $\Delta S_{A}\geq 0$, because by definition it is equal to the relative entropy between $\rho_A$ and $\rho_{A,Q}$, 
$\Delta S_{A}=\mathrm{Tr}[\rho_A(\log\rho_A-\log\rho_{A,Q})]$, which is actually non-negative~\cite{nielsen}; 
$(ii)$ $\Delta S_{A}=0$ if and only if the state is symmetric since, in this case, $\rho_A$ is block diagonal
in the eigenbasis of $Q_A$ and 
$\rho_{A, Q}=\rho_A$.

{\bf Results}

\paragraph{{\it A replica construction}.} The entanglement asymmetry  can be computed from the moments of 
the density matrices $\rho_A$ and $\rho_{A, Q}$ by exploiting the replica trick~\cite{holzhey94, cc-04}. 
Indeed, simply defining the R\'enyi entanglement asymmetry as 
\begin{equation}\label{eq:replicatrick}
  \Delta S_{A}^{(n)}=\frac{1}{1-n}\left[\log\mathrm{Tr}(\rho_{A, Q}^n)-\log\mathrm{Tr}(\rho_A^n)\right],
\end{equation}
one has that $\displaystyle{\lim_{n\to 1}}\Delta S_{A}^{(n)}=\Delta S_{A}$. 
As usual, the advantage of this construction is that, for integer $n$, $\Delta S_{A}^{(n)}$ can be accessed from (charged) partition  functions. 
Using the Fourier representation of the projector $\Pi_q$, the post-measurement density 
matrix $\rho_{A, Q}$ can be alternatively written in the form
\begin{equation}
 \rho_{A, Q}=\int_{-\pi}^\pi \frac{{\rm d}\alpha}{2\pi}e^{-i\alpha Q_A}\rho_A e^{i\alpha Q_A},
\end{equation}
and its moments as 
\begin{equation}\label{eq:FT}
 \mathrm{Tr}(\rho_{A, Q}^n)=\int_{-\pi}^\pi \frac{{\rm d}\alpha_1\dots{\rm d}\alpha_n}{(2\pi)^n} Z_n(\boldsymbol{\alpha}),
 \end{equation}
where $\boldsymbol{\alpha}=\{\alpha_1,\dots,\alpha_n\}$ and
\begin{equation}\label{eq:Znalpha}
 Z_n(\boldsymbol{\alpha})=
 \mathrm{Tr}\left[\prod_{j=1}^n\rho_A e^{i\alpha_{j,j+1}Q_A}\right],
\end{equation}
with $\alpha_{ij}\equiv\alpha_i-\alpha_j$ and $\alpha_{n+1}=\alpha_1$. 
Notice that, if $[\rho_A, Q_A]=0$, then $Z_n(\boldsymbol{\alpha})=Z_n(\boldsymbol{0})$, which implies $\mathrm{Tr}(\rho_{A, Q}^n)=\mathrm{Tr}(\rho_A^n)$ and $\Delta S_{A}^{(n)}=0$. 
Furthermore the order of terms in Eq. \eqref{eq:Znalpha} matters because $[\rho_A,Q_A]\neq0$.
We will refer to $Z_n(\boldsymbol{\alpha})$ as {\it charged moments} because they are a modification of the similar quantities 
introduced for the symmetry resolution of entanglement~\cite{goldstein}.

\begin{figure}
     \centering
     \includegraphics[width=\linewidth]{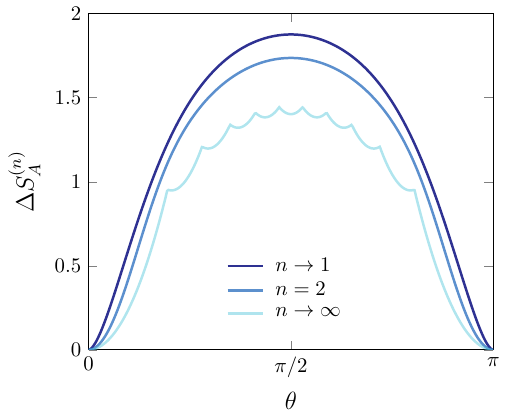}
     \caption{{\bf R\'enyi entanglement asymmetry $\Delta S_{A}^{(n)}$ 
     for the tilted ferromagnetic state.} We plot the analytic expression of Eq.~\eqref{eq:DeltaSnogaussian_exact} for this state
     as a function of the tilting angle $\theta$ for different values of  the replica index $n$ and subsystem size $\ell=10$.}
     \label{fig:tilted}
\end{figure}

\paragraph{{\it Tilted Ferromagnet}.} As warm up, 
we start with an undergraduate exercise. We consider 
an infinite spin chain prepared in the tilted ferromagnetic state,
i.e. the spins are not aligned with the quantization axis $z$,
\begin{equation}\label{eq:tilted_ferro}
 |\theta; \nearrow\nearrow\cdots\rangle=e^{-i\frac{\theta}{2} \sum_{j}\sigma_j^y}|\uparrow\uparrow\cdots\rangle.
\end{equation}
For $\theta\neq \pi m$, $m\in\mathbb{Z}$, this state breaks the $U(1)$ symmetry associated to the conservation
of the total transverse magnetization $Q=\frac{1}{2}\sum_j\sigma_j^z$. When $\theta=\pi m$, it 
corresponds to a fully polarized state in the $z$-direction, for which the transverse
magnetization is preserved. The angle $\theta$ controls how much the state breaks this
symmetry and, therefore, the state \eqref{eq:tilted_ferro} is an ideal testbed for the entanglement asymmetry, although it is a trivial product state.
Let the subsystem $A$ consist of $\ell$ contiguous sites of the chain; then $\Delta S_{A}=0$ for $\theta=\pi m$ and $\Delta S_{A}>0$ otherwise. 
Since the state is separable, $\mathrm{Tr}(\rho_A^n)=1$, and $Z_n(\boldsymbol{\alpha})$ is straightforwardly obtained as
\begin{equation}\label{eq:moments_tilted}
    Z_n(\boldsymbol{\alpha})=\prod_{j=1}^n \left[i\cos(\theta)\sin\left(\frac{\alpha_{j,j+1}}{2}\right)+\cos\left( \frac{\alpha_{j,j+1}}{2}\right)\right]^{\ell}.
\end{equation}
Plugging Eq. \eqref{eq:moments_tilted}  into the Fourier transform~\eqref{eq:FT}, we obtain 
\begin{equation}\label{eq:DeltaSnogaussian_exact}
 \Delta S_{A}^{(n)}=\frac{1}{1-n}\log\left[\cos^{2n\ell}\Big(\frac{\theta}{2}\Big)\sum_{p=0}^\ell\binom{\ell}{p}^n
 \tan^{2np}\Big(\frac{\theta}{2}\Big)\right].
\end{equation}
In Fig.~\ref{fig:tilted}, we plot this entanglement asymmetry as a function of $\theta\in[0,\pi]$. As expected, 
it vanishes for $\theta=0,\pi$ 
while it takes the maximum value at $\theta=\pi/2$, when all the spins point in the $x$ direction and the symmetry is maximally broken. 
Between these extremal points, $\Delta S_{A}$ is a monotonic function of $\theta$ (but this is not true for all $n$). 
For a large interval, it behaves as 
\begin{equation}\label{eq:DeltaSnogaussian}
 \Delta S_{A}^{(n)}=\frac{1}{2}\log \ell+\frac{1}{2}\log\frac{\pi n^{\frac{1}{n-1}}\sin^2\theta}{2}+O(\ell^{-1}).
\end{equation}
The limit $\theta \to 0$ 
is not well defined in Eq.~\eqref{eq:DeltaSnogaussian}. Indeed, the limits $\ell \to \infty$ and $\theta \to 0$ do not commute: to recover the symmetry, one should take first $\theta\to 0$ in Eq.~\eqref{eq:moments_tilted} and then consider the large interval regime.

\paragraph{{\it Quench to the XX spin chain}.} 
We now analyze the time evolution of the entanglement asymmetry after a quantum quench. 
We prepare the infinite spin chain in the state
\begin{equation}\label{eq:initial_state}
 \ket{\Psi(0)}=\frac{\ket{\theta;\nearrow\nearrow\cdots}-
 \ket{-\theta;\nearrow\nearrow\cdots}}{\sqrt{2}},
\end{equation}
which is the {\it cat} version of the symmetry-breaking state in Eq. \eqref{eq:tilted_ferro}. 
We then let it evolve
\begin{equation}\label{eq:quench}
 |\Psi(t)\rangle = e^{-it H} |\Psi(0)\rangle,
\end{equation}
with the symmetric XX Hamiltonian  ($[H, Q]=0$)
\begin{equation}
 H=-\frac{1}{4}\sum_{j=-\infty}^{\infty}\left[\sigma_{j}^x\sigma_{j+1}^x
 +\sigma_{j}^y\sigma_{j+1}^y\right].
\end{equation}
This Hamiltonian is diagonalized via the Jordan-Wigner transformation to fermionic operators
followed by a Fourier transform to momentum space~\cite{lms-61}.
The one-particle dispersion relation is $\epsilon(k)=-\cos(k)$.

\begin{figure}
     \centering
     \includegraphics[width=\linewidth]{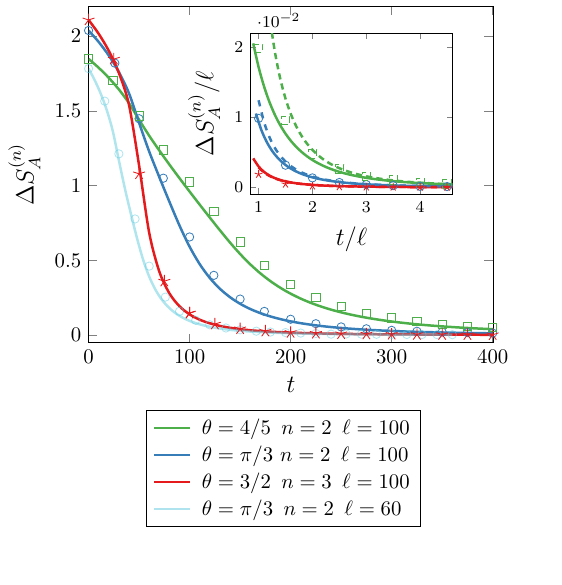}
     \caption{
     {\bf Time evolution of the R\'enyi entanglement asymmetry $\Delta S_{A}^{(n)}(t)$ after the quench~\eqref{eq:quench}.} 
     The symbols are the exact numerical results for various values of the subsystem length $\ell$, the replica index $n$, and the initial tilting angle $\theta$ (see Methods). 
     The continuous lines are our prediction obtained by plugging the charged moments reported in Methods into \eqref{eq:FT} and \eqref{eq:replicatrick}. 
     In the inset, we check the asymptotic behavior \eqref{eq:asympt} (full lines) and \eqref{exp2} (dashed) of $ \Delta S_{A}^{(n)}(t)$ for large $t/\ell$.}
     \label{fig:delta_S}
\end{figure}

\paragraph{\textit{The entanglement asymmetry after the quench}.}
At time $t=0$, the entanglement asymmetry behaves asymptotically as Eq.~\eqref{eq:DeltaSnogaussian}; for $t>0$,
$\Delta S_{A}^{(n)}(t)$ is analytically derived in Methods by adapting the {\it quasiparticle picture} of entanglement dynamics~\cite{cc-05, ac-17, ac-18} to the charged 
moments~\eqref{eq:Znalpha} and then taking the Fourier transform~\eqref{eq:FT}.
The resulting curves are plotted in Fig.~\ref{fig:delta_S} as a function of $\zeta=t/\ell$ for several values of $\theta$, finding a remarkable agreement with the exact numerical values (symbols).
We can also write a very effective closed-form approximation of $\Delta S^{(n)}_{A}(t)$,  
\begin{equation}\label{eq:asympt}
\begin{split}
    \Delta S_{A}^{(n)}(t)\,\simeq \,& \frac{\pi^2 b(\zeta)\ell}{24},\\
    b(\zeta)\,=\,&\frac{\sin^2\theta}{2}-\int_0^{2\pi} \frac{\mathrm{d}k}{2\pi}\mathrm{min}(2\zeta |\epsilon'(k)|,1)\sin^2\Delta_k,\end{split}
\end{equation}
which is independent of the replica index $n$ (see Methods for the definition of $\Delta_k$). 
This approximation becomes {\it exact} in the limit of large $\zeta$ and its effectiveness, also for not too large $\zeta$, is proven by the inset of Fig. \ref{fig:delta_S}.

We now discuss some relevant features of the entanglement asymmetry and show that it encodes a lot of new physics.
First, as expected \cite{FCEC14,pvc-16}, $\Delta S_{A}^{(n)}(t)$ tends to zero for large $\zeta$ (i.e. large $t$) and the $U(1)$ symmetry, broken by the initial state, is restored. 
This is analytically shown by Eq. \eqref{eq:asympt} that indeed at leading order in large $\zeta$ is 
\begin{equation}
    \Delta S_{A}^{(n)}(t)\simeq \frac{\pi}{1152} \Big(1+8\frac{\cos^2\theta}{\sin^4\theta} \Big) \frac{\ell}{\zeta^3}\,,
    \label{exp2}
\end{equation}
i.e. it vanishes for large times as $t^{-3}$ for any value of $\theta$.
This decay is determined by the quasiparticles with the slowest velocity $|\epsilon'(k)|$,
which in this case are those with momentum around $k=0$ and $\pi$.
Another characteristic, following from having a space-time scaling, is that larger subsystems require more time to recover the symmetry, as it is 
clear from Fig.~\ref{fig:delta_S} and Eq. \eqref{eq:asympt}:
this justifies the significance of the definition of $\Delta S_{A}^{(n)}$ in terms of $\rho_A$ rather than the full state $\ket{\Psi}$. 
Finally, a very odd and intriguing feature is that the more the symmetry is initially broken, i.e. the larger $\theta$, the smaller the time to restore it.     
This is a quantum Mpemba effect \cite{mpe}: more the system is out of equilibrium, the faster it relaxes.
At a qualitative level this is a consequence of the fact that for larger symmetry breaking there is a sharper drop of the (entanglement) asymmetry  
at short time, see Fig. \ref{fig:delta_S}, before the truly asymptotic behavior takes place. 
Furthermore, we can quantitatively understand the quantum Mpemba effect: 
from Eq. \eqref{exp2} the prefactor of the $t^{-3}$ decay is a monotonously decreasing function of $\theta$ in $[0,\pi/2]$. 
Thus the quantum Mpemba effect is not as controversial as its classical version \cite{mpemba}.
To the best of our knowledge this awkward effect was not known in the literature, showing the power of the entanglement  asymmetry to identify new physics.

\paragraph{Quantum Mpemba effect.} The quantum Mpemba effect is not a prerogative of integrable free systems, such as the XX spin chain, but it turns out to be much more general and robust. To show this, we analyze now a global quantum quench having as initial state the tilted ferromagnetic configuration of Eq. \eqref{eq:tilted_ferro} and evolving with the interacting Hamiltonian 
\begin{multline}\label{eq:Hamiltonian2}
 H=-\frac{1}{4}\sum_{j=1}^{N}\left[\sigma_{j}^x\sigma_{j+1}^x
+\sigma_{j}^y\sigma_{j+1}^y+\Delta \sigma_{j}^z\sigma_{j+1}^z\right]\\
-\frac{J_2}{4}\sum_{j=1}^{N}\left[\sigma_{j}^x\sigma_{j+2}^x
+\sigma_{j}^y\sigma_{j+2}^y+\Delta_2 \sigma_{j}^z\sigma_{j+2}^z\right],
\end{multline}
where $N$ is the total number of spins. 
This Hamiltonian commutes with the transverse magnetization $Q=\frac12\sum_j \sigma_j^z$. 
For $J_2=0$, it corresponds to the Heisenberg XXZ
spin chain with anisotropy parameter $\Delta$, which is the prototype
of all interacting integrable models. For $\Delta=J_2=0$, we
recover the XX spin chain of the previous paragraphs. 
For $J_2\neq 0$, the next nearest neighbor couplings break integrability~\cite{non-integrable}.

The $U(1)$ symmetry is expected to be restored after a generic quench to the Hamiltonian \eqref{eq:Hamiltonian2}~\cite{FCEC14}. 
In fact, at late times, the local stationary behavior is described by a statistical ensemble, 
corresponding to thermal or generalized Gibbs for chaotic or integrable systems respectively~\cite{gibbs,gibbs2,gibbs3,gge,gge2}. 
In one dimensional quantum systems, the Mermin-Wagner theorem forbids the spontaneous breaking of a continuous symmetry at finite temperature.
In the quench, the finite energy density of the initial state plays the role of an effective temperature, causing in general symmetry restoration 
(with the exception of very few pathological cases).

\begin{figure}[t]
     \centering    
     \includegraphics[width=0.49\textwidth]{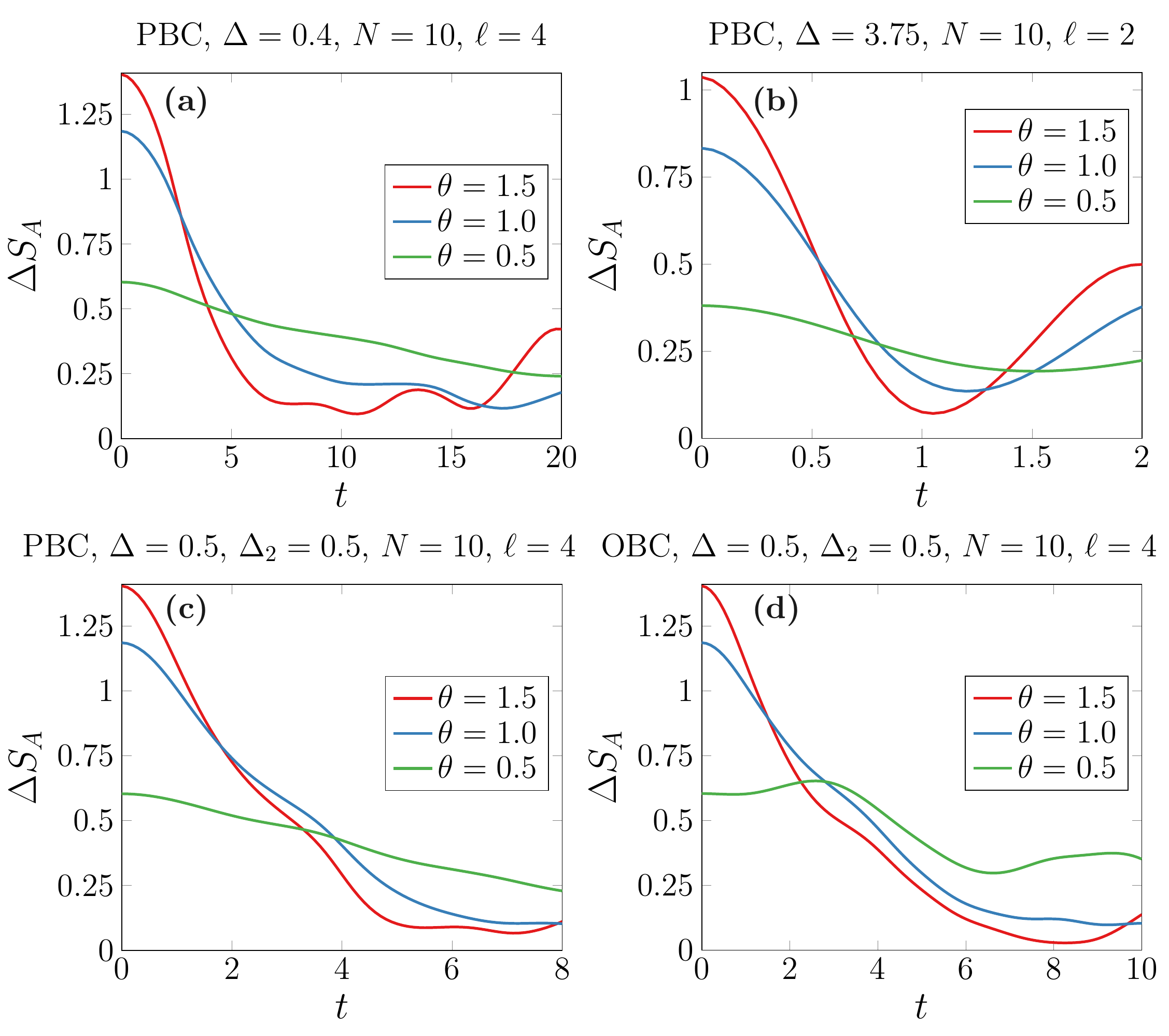}
     \caption{{\bf Quantum Mpemba effect in interacting integrable and non-integrable spin chains.} We plot the time evolution of the entanglement asymmetry $\Delta S_{A}(t)$ after preparing the spin chain at $t=0$ in the tilted ferromagnetic state 
     of Eq.~\eqref{eq:tilted_ferro} and performing a sudden quench to the Hamiltonian $H$ given in Eq. \eqref{eq:Hamiltonian2} with total length $N=10$. 
     The continuous lines have been obtained via exact diagonalization for different choices of the subsystem length $\ell$, initial tilting angle $\theta$, and of the couplings and 
     boundary conditions in the evolution Hamiltonian $H$. In panels {\bf a} and {\bf b}, $J_2=0$, and $H$ corresponds to the XXZ spin chain with anisotropy parameter $\Delta$; in both cases, we take PBC for the chain. In panels {\bf c} and {\bf d}, $J_2=1$, and the chain is non-integrable; in panel {\bf c}, we consider PBC while in panel {\bf d} we choose
     OBC with the subsystem $A$ placed in the 
     middle of the chain.}

     \label{fig:delta_S2}
\end{figure}

In Fig.~\ref{fig:delta_S2}, we plot the time evolution of $\Delta S_A$
after a quench using the Hamiltonian~\eqref{eq:Hamiltonian2} with $N=10$ spins
for different values of the couplings and initial tilting angle $\theta$. In all the cases, the curves have been obtained by applying exact diagonalization. 
In panels {\bf a} and {\bf b} of Fig.~\ref{fig:delta_S2}, we perform a quench to a periodic XXZ chain ($J_2=0$) with interaction $\Delta=0.4$ (panel {\bf a}) and $\Delta=3.75$ (panel {\bf b}). In panels {\bf c} and {\bf d} of Fig.~\ref{fig:delta_S2}, the post-quench Hamiltonian contains next nearest neighbor terms ($J_2=1$) and, therefore, is non-integrable. Panel {\bf c} corresponds to periodic boundary conditions (PBC) while
in panel {\bf d} we consider open boundary conditions (OBC)
with the subsystem located at the middle of the chain.
In all the plots, the quantum Mpemba effect is clearly visible: the more the symmetry is initially broken, 
the faster $\Delta S_A(t)$ decays to zero after the quench;
this is true, although the finite size of the system causes revivals that prevent us
from observing the restoration in a neat way as happens in the thermodynamic limit in 
Fig.~\ref{fig:delta_S}.

In conclusion, Fig.~\ref{fig:delta_S2} shows that quantum Mpemba effect occurs 
under very general conditions (both for integrable and non-integrable
interactions with different boundary conditions), even for (sub)systems of few sites, which makes possible
to observe it experimentally in, e.g., ion trap setups. 

{\bf Discussion}

In this work, we introduced the entanglement asymmetry, a probe to study how much a symmetry is broken at the level of subsystems of many-body systems. 
As an application to show its potential, we have studied its dynamics after a quench from an initial state breaking a $U(1)$ symmetry and evolving with a Hamiltonian preserving it. 
We showed that the entanglement asymmetry detects neatly all the physical relevant features of the dynamics and in particular the restoration of the symmetry at late times. 
It also identifies the appearance of an unexpected Mpemba effect, a phenomenon that, as we have seen, happens in many settings that can be studied through the entanglement asymmetry.
It is then very important to study other quench protocols (e.g. different initial state, interacting Hamiltonians, etc.) 
and understand how to modify the quasiparticle description, following e.g. Ref.~\cite{bkalc-22}, to describe these more general situations. \\
\indent We can easily imagine many other applications of the entanglement asymmetry. 
The first one is in equilibrium situations that have been left out here. 
In this respect, it would be useful to recast the charged moments~\eqref{eq:Znalpha} in terms of twist fields~\cite{cc-04, cdca-07}  within the path-integral approach: 
this would allow us to explore more complicated situations, e.g. the symmetry breaking from $SU(2)$ to $U(1)$, which are also relevant in high-energy physics \cite{chmp-20}.
Similarly, our setup can be extended  to non-Abelian symmetries~\cite{cdm-21} to explore, e.g., how the asymptotic behavior
of $\Delta S_A$ with the subsystem size of Eq.~\eqref{eq:DeltaSnogaussian} depends on the symmetry group. \\ \indent Finally, $\Delta S_{A}^{(n)}(t)$, with $n$ integer $n\geq 2$, can be experimentally accessible by developing a protocol based on the random measurement toolbox \cite{Vermersch2019scrambling,RMtoolbox,shadows}. 
This would require the post-selection of data from an experiment like the one in \cite{brydges-2019}, but with an initial state breaking the $U(1)$ symmetry.

{\bf Methods}

We provide here the details about the derivation of the numerical and analytical results reported in the Results section.

{\it Numerical techniques.} We choose as initial state the linear combination of Eq.~\eqref{eq:initial_state}, instead of
Eq.~\eqref{eq:tilted_ferro}, because, after a Jordan-Wigner transformation, the corresponding reduced density matrix is Gaussian in terms of the fermionic operators $\boldsymbol{c}_j=(c_j^\dagger, c_j)$.  
We can then use Wick theorem to express $\rho_A(t)$  in terms of the two-point correlation matrix 
\begin{equation}
 \Gamma_{jj'}(t)=2\Tr\left[\rho_A(t) \boldsymbol{c}_j^\dagger
 \boldsymbol{c}_{j'}\right]-\delta_{jj'},
\end{equation}
with $j,j'\in A$ \cite{p-03}. If $A$ is a subsystem of length $\ell$, then $\Gamma(t)$
has dimension $2\ell\times 2\ell$ and entries \cite{FC08}
\begin{equation}\label{eq:Gamma}
\Gamma_{jj'}(t)=\int_{0}^{2\pi}\frac{{\rm d}k}{2\pi}\mathcal{G}(k, t)e^{-ik(j-j')},
\end{equation}
with
\begin{multline}
    \mathcal{G}(k,t)=\left(\begin{array}{cc}
    \cos\Delta_k & -i e^{-2it\epsilon(k)}\sin\Delta_k\\
     ie^{2it\epsilon(k)}\sin\Delta_k& -\cos\Delta_k
    \end{array}\right),\\
 \cos\Delta_k=\frac{2\cos(\theta)-(1+\cos^2\theta)\cos(k)}{1+\cos^2\theta-2\cos(\theta)\cos(k)}.
\end{multline}
Under the Jordan-Wigner transformation, the transverse magnetization
is mapped to the fermion number operator and  $e^{i\alpha Q_A}$ turns out to be Gaussian, too.
Therefore, $ Z_n(\boldsymbol{\alpha})$ in Eq.~\eqref{eq:Znalpha} is the trace of the product of  Gaussian fermionic operators, $\rho_A$ and $e^{i\alpha_{j,j+1} Q_A}$. 
Employing their composition properties~\cite{bb-69,FC10}, we express $ Z_n(\boldsymbol{\alpha})$ as a determinant involving the corresponding correlation matrices, finding
\begin{equation}\label{eq:numerics}
  Z_n(\boldsymbol{\alpha}, t)=\sqrt{\det\left[\left(\frac{I-\Gamma(t)}{2}\right)^n
  \left(I+\prod_{j=1}^n W_j(t)\right)\right]},
\end{equation}
with $W_j(t)=(I+\Gamma(t))(I-\Gamma(t))^{-1}e^{i\alpha_{j,j+1} n_A}$ and $n_A$ is a diagonal matrix with $(n_A)_{2j,2j}=1$, $(n_A)_{2j-1,2j-1}=-1$, $j=1, \cdots, \ell$. 
We use Eq.~\eqref{eq:numerics} to numerically compute the time evolution of the R\'enyi entanglement asymmetry $\Delta S^{(n)}_{A}(t)$ in Fig.~\ref{fig:delta_S}
and test the analytical predictions presented in this work.

\begin{figure}[h]
     \centering
     \includegraphics[width=0.45\textwidth]{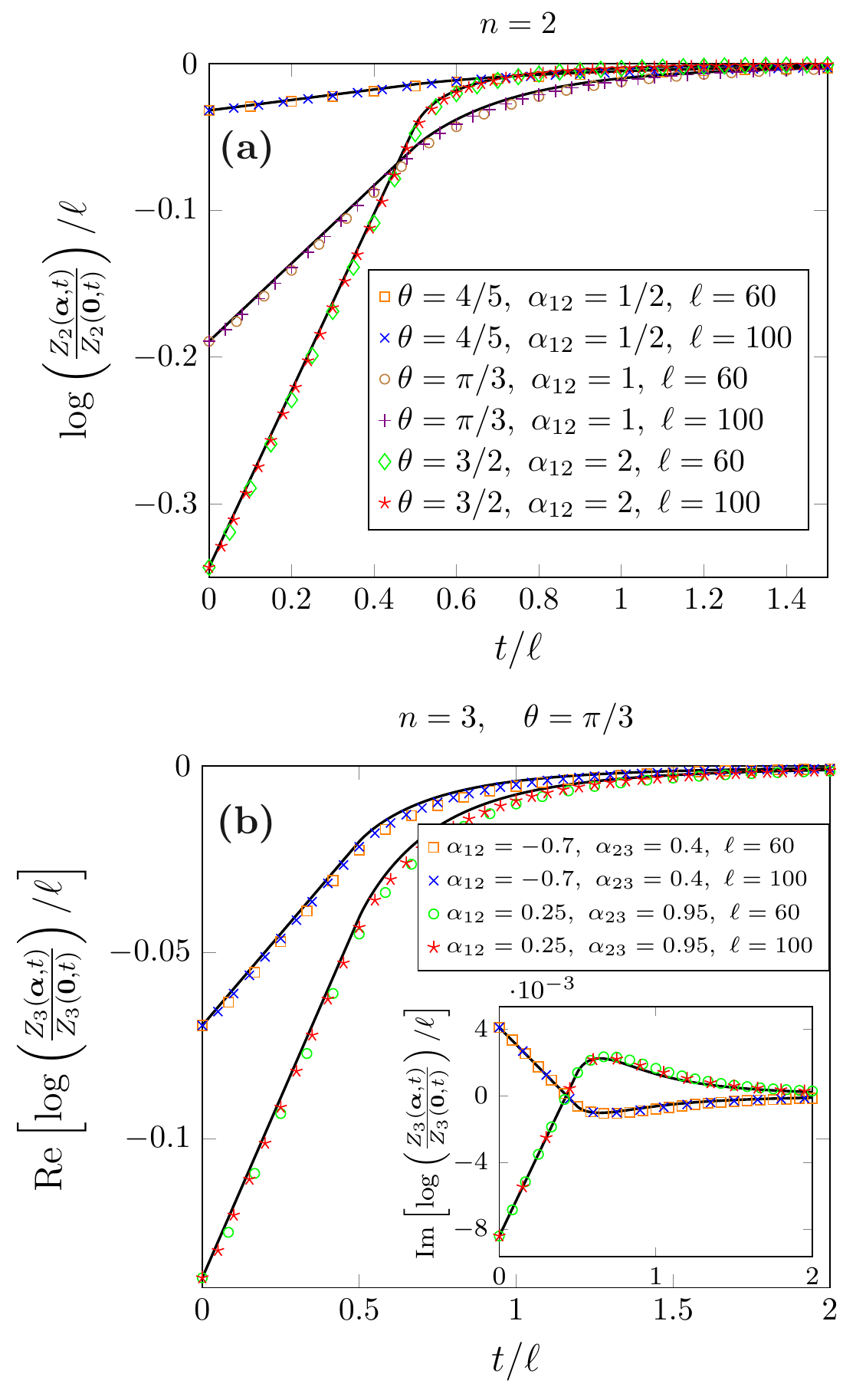}
     \caption{{\bf Time evolution of the charged moments $Z_n(\boldsymbol{\alpha},t)$ after the 
     quench~\eqref{eq:quench}.} We plot them as a function of $t/\ell$ for the replica indices $n=2$ (panel {\bf a}) and $n=3$ (panel {\bf b}) and several
values of the initial tilting angle $\theta$, the subsystem size $\ell$, and the phases $\alpha_{j,j+1}$. The symbols 
were obtained numerically using Eq.~\eqref{eq:numerics} and the continuous lines correspond to the analytic prediction~\eqref{eq:z2alpha}.}
     \label{fig:moments}
\end{figure}

{\it Analytic computation.} After the quench, the natural ballistic regime is the scaling limit $t, \ell \to \infty$ with $\zeta=t/\ell$ fixed \cite{FC08,cef-12I}, in which we find
\begin{equation}\label{eq:z2alpha}
    Z_n(\boldsymbol{\alpha}, t)=Z_n(\boldsymbol{0}, t)e^{\ell(A(\boldsymbol{\alpha})+B(\boldsymbol{\alpha},\zeta))},
\end{equation}
where the functions $A(\boldsymbol{\alpha})$ and $B(\boldsymbol{\alpha},\zeta)$ read, respectively,
\begin{equation}\label{eq:Balpha}
\begin{split}
    A(\boldsymbol{\alpha})=&\displaystyle \int_0^{2\pi}\frac{\mathrm{d}k}{2\pi}\log\prod_{j=1}^nf(e^{i \Delta_{k}},\alpha_{j,j+1}),\\
    B(\boldsymbol{\alpha}, \zeta)=&-\int_0^{2\pi}\frac{\mathrm{d}k}{2\pi}\mathrm{min}(2\zeta |\epsilon'(k)|,1)\\&\times\log\prod_{j=1}^n f(e^{i\Delta_k},\alpha_{j,j+1}),
    \end{split}
\end{equation}
and $f(\lambda,\alpha)$ is defined as
\begin{equation}
   f(\lambda,\alpha)=i\lambda \sin\left(\frac{\alpha}{2}\right)+\cos\left( \frac{\alpha}{2}\right).
\end{equation}
Notice that in Eq.~\eqref{eq:Balpha} there is a factorization in the replica space indexed by $j$.
This cumbersome expression does not come out of a magician hat, but from the  quasiparticle picture~\cite{cc-05, ac-17, ac-18}:
the time evolution of the entanglement is given by the pairs of entangled excitations shared by
$A$ and $B$ that are created after the quench and propagate ballistically with momentum $\pm k$.
Let  us explain how to apply this idea to deduce  Eq.~\eqref{eq:Balpha}.
According to Refs.~\cite{FCEC14,pvc-16}, in the quench protocol analyzed here, the $U(1)$ symmetry is restored in the large time limit, i.e. $\Delta S_{A}^{(n)}(t) \to 0$. 
Therefore, $Z_n(\boldsymbol{\alpha},t)$ has to tend to $ Z_n(\boldsymbol{0},t)$, which implies $B(\boldsymbol{\alpha}, \zeta)\to -A(\boldsymbol{\alpha})$ as $\zeta\to\infty$. 
At time $t=0$, plugging the initial state of Eq.~\eqref{eq:initial_state} in the definition of the charged moments~\eqref{eq:Znalpha}, we obtain that, for large $\ell$, 
$Z_n(\boldsymbol{\alpha}, 0)\sim e^{A(\boldsymbol{\alpha})\ell}/2^{n-1}$ with
\begin{equation}\label{eq:Aalpha}
  A(\boldsymbol{\alpha})= \log\prod_{j=1}^n e^{i\sigma_j/2}f(\cos(\theta),\alpha_{j,j+1}-\sigma_j),
\end{equation}
where $\sigma_j=0$ if $|\alpha_{j,j+1}| \leq \pi/2$ and  $\sigma_j=\pi$ otherwise. 
Considering Eq.~\eqref{eq:Aalpha}, we notice that
$Z_n(\boldsymbol{\alpha},0)$ factorizes into 
\begin{equation}
    Z_n(\boldsymbol{\alpha},0)\sim 2\prod_{j=1}^n \frac{e^{i\sigma_j/2}}{2} \mathrm{Tr}(\rho_A(0) e^{i(\alpha_{j,j+1} -\sigma_j)Q_A}).
\end{equation}
The expectation value $\mathrm{Tr}(\rho_A(0) e^{i\alpha Q_A})$ is the full counting statistics (FCS) of the transverse magnetization in the subsystem $A$. 
We can now take advantage of the fact that $\ket{\Psi(0)}$ is also the ground state of a XY spin 
chain 
to exploit the knowledge of the FCS in that system~\cite{cherng07, ia-13, stephan14, ARV21, groha18} (the corresponding parameters $h, \gamma$ of the XY chain are given by $\gamma^2+h^2=1$ and $\cos^2 \theta=(1-\gamma)/(1+\gamma)$). In particular, employing the results of Ref.~\cite{ARV21}, we can rewrite $A(\boldsymbol{\alpha})$ in Eq.~\eqref{eq:Aalpha} as an integral in momentum space
\begin{equation}\label{eq:Balpha_infty}
A(\boldsymbol{\alpha})=-B(\boldsymbol{\alpha}, \zeta\to \infty)=\displaystyle \int_0^{2\pi}\frac{\mathrm{d}k}{2\pi}\log\prod_{j=1}^nf(e^{i \Delta_{k}},\alpha_{j,j+1}).
\end{equation}
Now, using the quasiparticle picture, the integrand in Eq.~\eqref{eq:Balpha_infty} can be interpreted as the contribution to $B(\boldsymbol{\alpha}, \zeta)$ from each entangled excitation of momentum $k$ created after the quench. Since they propagate with velocity $|\epsilon'(k)|$, the number of these pairs shared between $A$ and its complement at time $t$ is determined by $\mathrm{min}(2t |\epsilon'(k)|,\ell)$. Combining these two ingredients, we get Eq.~\eqref{eq:z2alpha}. 
This approach makes also clear the crucial role that entanglement plays in the restoration of the symmetry. Likely this expression can be rigorously derived by properly adapting the calculations for the symmetry resolved entanglement~\cite{pbc-21-1,pbc-21}, but this is far beyond the scope of this work.
In Fig.~\ref{fig:moments}, we check Eq.~\eqref{eq:z2alpha} against exact numerical computations performed using Eq.~\eqref{eq:numerics} for different values of $n$, $\theta$, and $\boldsymbol{\alpha}$, finding a remarkable agreement: note that Eq.~\eqref{eq:z2alpha} is exact 
for $\ell\to\infty$ and the points are closer to the curves for larger $\ell$.
Finally, when in Eq.~\eqref{eq:z2alpha} $A(\boldsymbol{\alpha})+B(\boldsymbol{\alpha},\zeta)$ is close to zero, the Fourier transform~\eqref{eq:FT} can be done analytically and we obtain the approximation for the entanglement asymmetry in Eq.~\eqref{eq:asympt}.

{\bf Acknowledgements}  

We thank Jerome Dubail, Viktor Eisler, Maurizio Fagotti, Israel Klich, Lorenzo Piroli, Eric Vernier, and Lenart Zadnik for useful discussions. 
All the authors acknowledge support from ERC under Consolidator grant number 771536 (NEMO). SM thanks
support from Caltech Institute for Quantum Information and Matter and the Walter Burke
Institute for Theoretical Physics at Caltech.









\end{document}